\documentclass[10pt,journal]{IEEEtran}

\usepackage[latin1,utf8]{inputenc} 
\usepackage{cite} 
\usepackage{microtype} 
\usepackage{color} 
\usepackage{graphicx} 
\usepackage{epstopdf} 
\usepackage{amsmath}
\usepackage{amssymb}
\usepackage{amsfonts}
\usepackage{amsthm} 
\usepackage{array} 
\usepackage[caption=false,font=footnotesize]{subfig} 
\usepackage{color} 
\usepackage{hyperref}
\usepackage{booktabs}
\usepackage[nolist]{acronym}
\usepackage{cleveref}
\usepackage{bm}
\usepackage{multirow}
\usepackage{placeins}
\usepackage{flushend}

\newcolumntype{?}{!{\vrule width 1pt}}

\definecolor{myBlue}{rgb}{0,0,0.55}
\hypersetup{
  linkcolor=  myBlue,
  citecolor=  myBlue,
  urlcolor=   myBlue,
  pdffitwindow=       true,
  breaklinks=         true,
  pdfcenterwindow=    true,
  pdfstartview=       FitH,
  bookmarks=          false,
  bookmarksopen=      false,   
  bookmarksnumbered=  false,   
  pdfhighlight=       /P,
  linktocpage=        true,   
  colorlinks=         false,
  pdfborder=          0 0 0,
  pdftitle=           {Recursive CSI Quantization of Correlated MIMO Channels Supporting Deep Learning Classification},
  pdfauthor=          {Stefan Schwarz},
  pdfkeywords=        {},
	pdfproducer = 			{},
	pdfcreator =				{}
  }

\usepackage[switch]{lineno}
\newcommand*\patchAmsMathEnvironmentForLineno[1]{%
 \expandafter\let\csname old#1\expandafter\endcsname\csname #1\endcsname
 \expandafter\let\csname oldend#1\expandafter\endcsname\csname end#1\endcsname
 \renewenvironment{#1}%
    {\linenomath\csname old#1\endcsname}%
    {\csname oldend#1\endcsname\endlinenomath}}%
\newcommand*\patchBothAmsMathEnvironmentsForLineno[1]{%
 \patchAmsMathEnvironmentForLineno{#1}%
 \patchAmsMathEnvironmentForLineno{#1*}}%
\AtBeginDocument{%
\patchBothAmsMathEnvironmentsForLineno{equation}%
\patchBothAmsMathEnvironmentsForLineno{align}%
\patchBothAmsMathEnvironmentsForLineno{flalign}%
\patchBothAmsMathEnvironmentsForLineno{alignat}%
\patchBothAmsMathEnvironmentsForLineno{gather}%
\patchBothAmsMathEnvironmentsForLineno{multline}%
}

\makeatletter
\renewcommand{\maketag@@@}[1]{\hbox{\m@th\normalsize\normalfont#1}}%
\makeatother

\newcommand{\EEb}[1]{\mathbb{E}\left(#1\right)}

\newcommand{\fro}[1]{\left\|#1\right\|}
\newcommand{\HH}[1]{#1^{\mathsf{H}}}
\newcommand{\HHb}[1]{\left(#1\right)^{\mathsf{H}}}
\newcommand{\mbf}[1]{\mathbf{#1}}
\newcommand{\mbfh}[1]{\HH{\mbf{#1}}}
\newcommand{\vect}[1]{\mathrm{vec}\left(#1\right)}

\newcommand{\hmbf}[1]{\bm\hat{\mathbf{#1}}}

\newcommand{\argmin}{\operatornamewithlimits{arg\,min}}
\newcommand{\argmax}{\operatornamewithlimits{arg\,max}}

\newcommand{\chord}[1]{\mathrm{d_c^2}\left(#1\right)}

\newcommand{\trace}[1]{\mathrm{tr}\left(#1\right)}

\newcommand{\spans}[1]{\mathrm{span}\left(#1\right)}

\newcommand{\GG}[1]{\mathcal{G}\left(#1\right)}

\crefname{figure}{Fig.}{Figs.}
\crefname{equation}{Eq.}{Eqs.}
\crefname{theorem}{Theorem}{Theorems}

\begin{acronym}
\acro{3GPP}{Third generation partnership project}
\acro{ABF}{transmit beamformed Alamouti space-time/frequency coding}
\acro{BD}{block-diagonalization}
\acro{BF}{beamforming}
\acro{CDF}{cumulative distribution function}
\acro{CSI}{channel state information}
\acro{CSIT}{CSI at the transmitter}
\acro{DNN}{deep neural network}
\acro{DoF}{degrees of freedom}
\acro{eMBMS}{enhanced multimedia broadcast/multicast service}
\acro{FDD}{frequency division duplex}
\acro{IA}{interference-alignment}
\acro{iid}[i.i.d.]{independent and identically distributed}
\acro{LBM}{leakage-based multicasting}
\acro{LTE}{long term evolution}
\acro{MC-IC}{multicast interference channel}
\acro{MISO}{multiple-input single-output}
\acro{MIMO}{multiple-input multiple-output}
\acro{MSE}{mean-squared error}
\acro{RBD}{regularized block-diagonalization}
\acro{RVQ}{random vector quantization}
\acro{SDR}{semidefinite relaxation}
\acro{SINR}{signal to interference and noise ratio}
\acro{SNR}{signal to noise ratio}
\acro{SQBC}{subspace quantization based combining}
\acro{SVD}{singular value decomposition}
\acro{SVR}{singular value ratio}
\acro{UMTS}{universal mobile telecommunications system}
\end{acronym}

\begin{document}
\title{Recursive CSI Quantization of Time-Correlated MIMO Channels by Deep Learning Classification}

\author{Stefan Schwarz, \textit{Senior Member, IEEE} \\
\vspace*{-23.5pt}
\thanks{
\indent S. Schwarz is with the Institute of Telecommunications, TU Wien, Austria; email: sschwarz@nt.tuwien.ac.at; tel.: +43-58801-38985. He leads the Christian Doppler Laboratory for Dependable Wireless Connectivity for the Society in Motion. The financial support by the Austrian Federal Ministry for Digital and Economic Affairs and the National Foundation for Research, Technology and Development is gratefully acknowledged.}
}

\maketitle

\begin{abstract} 
In \ac{FDD} \ac{MIMO} wireless communications, limited \ac{CSI} feedback is a central tool to support advanced single- and multi-user \ac{MIMO} beamforming/precoding. To achieve a given \ac{CSI} quality, the \ac{CSI} quantization codebook size has to grow exponentially with the number of antennas, leading to quantization complexity, as well as, feedback overhead issues for larger \ac{MIMO} systems. We have recently proposed a multi-stage recursive Grassmannian quantizer that enables a significant complexity reduction of \ac{CSI} quantization. In this paper, we show that this recursive quantizer can effectively be combined with deep learning classification to further reduce the complexity, and that it can exploit temporal channel correlations to reduce the \ac{CSI} feedback overhead.  
\end{abstract}
\vspace*{-5pt}
\acresetall 

\vspace*{-10pt}
\section{Introduction}
\label{sec:Introduction}

Limited \ac{CSI} feedback is a well-established technique for supporting efficient \ac{MIMO} transmissions in \ac{FDD} systems~\cite{Love2008,Choi2013,Park2016,Kwon2020}. Often, the framework of Grassmannian \ac{CSI} quantization is adopted, since subspace information is required for many popular transmit precoding schemes. A large number of different ways for constructing Grassmannian quantization codebooks exists; e.g.,~\cite{Medra15,Decurninge2017,Ratnam18,Schwarz_JSTSP19} to mention just a few of the more recent constructions. 

Generally, in case of memoryless quantization of isotropic channels, such as, \ac{iid} Rayleigh fading channels, it is known that maximally spaced subspace packings achieve optimal quantization performance in terms of subspace chordal distance; however, such packings are difficult to construct for larger \ac{MIMO} systems and codebook sizes~\cite{Dhillon2007,Laue2017,Tahir_SPL19}. When adopting Grassmannian quantization in larger-scale \ac{MIMO} systems and/or for high resolution quantization, one faces two main challenges: 1) quantization complexity and 2) feedback overhead. The former issue can effectively be tackled, if the channel exhibits structure that can be exploited for quantization; e.g., in the millimeter wave band, the channel is often assumed to be sparse, which allows for efficient parametric \ac{CSI} quantization by sparse decomposition~\cite{Luo2017,Xie2017,Schwarz_JWCN2018}. Such techniques, however, are not applicable in \ac{iid} Rayleigh fading situations. Also, recently, a number of approaches that utilize \acp{DNN} have been proposed to enable efficient \ac{CSI} quantization~\cite{Wang2019,Liu2019,Jang20}; yet, these publications mostly consider relatively low resolution quantization, as neural networks are hard to train for large quantization codebooks. When the channel exhibits temporal correlation, quantizers with memory, such as, differential quantizers or techniques based on recurrent neural networks, can provide significantly better performance than memoryless approaches~\cite{Inoue2009,Sacristan2010,Ayach2012,Schwarz-ACSQ2013/1,Schwarz_SPL_2014,Ge2018,Lu2019,Li2020}. Yet, they mostly require adaptation of the quantization codebook on the fly or online neural network learning, which can be prohibitive in terms of complexity.

\emph{Contribution}: In~\cite{Schwarz_SPL20}, we have proposed a recursive multi-stage quantization approach that can reduce the complexity of high resolution Grassmannian quantization in moderate to large-scale \ac{MIMO} systems by orders of magnitude. In this paper, we show that this approach can effectively be enhanced by \ac{DNN} classification to further reduce the implementation complexity and, thus, support high resolution Grassmannian quantization with low complexity. Hence, rather than adopting an end-to-end \ac{DNN} approach, we propose to enhance well-known model-based \ac{CSI} quantizers by neural network features. We furthermore propose a simple approach to exploit temporal channel correlation in recursive multi-stage quantization, by selectively updating the individual stages of the quantizer.

\emph{Notation}: The Grassmann manifold of $m$-dimensional subspaces of the $n$-dimensional Euclidean space is $\GG{n,m}$. The trace of matrix $\mbf{A}$ is $\trace{\mbf{A}}$, the conjugate-transpose is $\HH{\mbf{A}}$, the Frobenius norm is $\fro{\mbf{A}}$ and vectorization is $\vect{\mbf{A}}$. The $m$-dimensional subspace spanned by the columns of $\mbf{A} \in \mathbb{C}^{n \times m}$, $m\leq n$ is $\spans{\mbf{A}}$. The expected value of random variable $x$ is $\EEb{x}$. The operation $a_\text{min} = \argmin_{a \in \mathcal{A}} f(a)$ determines the minimizer $a_\text{min}$ of the function $f(a)$ over the set $\mathcal{A}$. The size of set $\mathcal{A}$ is $\left|\mathcal{A}\right|$. The vector-valued complex Gaussian distribution with mean $\boldsymbol{\mu}$ and covariance $\mbf{C}$ is $\mathcal{CN}\left(\boldsymbol{\mu},\mbf{C}\right)$. The zeroth-order Bessel function of the first kind is $J_0(\cdot)$. 

\vspace*{-5pt}
\section{Channel Model}
\label{sec:Channel}

We consider a \ac{MIMO} wireless communication system with $n$ transmit- and $m$ receive-antennas, where $n > m$. We denote the frequency-flat baseband \ac{MIMO} channel matrix at time instant $k$ as $\mbf{H}[k] \in \mathbb{C}^{n \times m}$. We consider a spatially uncorrelated Rayleigh fading channel, i.e., $\vect{\mbf{H}[k]} \sim \mathcal{CN}\left(\mbf{0},\mbf{I}_{n m}\right)$, as caused by a strong scattering environment.

We assume that the \ac{MIMO} channel follows a stationary Gaussian stochastic process with temporal auto-correlation function parametrized by the time-lag $\Delta k$ according to
\begin{gather}
	\boldsymbol{\Gamma}[\Delta k] = \mathbb{E}\left(\vect{\mbf{H}[k]} \HHb{\vect{\mbf{H}[k + \Delta k]}}\right).
\end{gather}
Considering, for example, Clarke's Doppler spectrum~\cite{Clarke1968}, the auto-correlation function is $\boldsymbol{\Gamma}[\Delta k] = J_0\left(2 \pi \nu_d \Delta k\right) \mbf{I}_{n m}$, where $\nu_d = f_d T_s$ is the normalized Doppler shift, $f_d$ is the maximal absolute Doppler shift and $T_s$ is the symbol time interval. 

\section{Grassmannian Quantization}
\label{sec:quant}

In this paper, we focus on Grassmannian \ac{CSI} quantization at the receiver, in order to provide \ac{CSI} feedback to the transmitter. For this purpose, we apply a compact size \ac{SVD} to the channel
\begin{gather}
	\mbf{H}[k] = \mbf{U}[k] \boldsymbol{\Sigma}[k] \HH{\mbf{V}[k]}, 
\end{gather}  
and utilize the orthogonal basis $\mbf{U}[k] \in \mathbb{C}^{n \times m}$, consisting of the left singular vectors corresponding to the non-zero singular values, as relevant \ac{CSI} to represent the $m$-dimensional subspace spanned by the channel $\spans{\mbf{H}[k]} = \spans{\mbf{U}[k]}$.

\subsection{Single-Stage Quantization}
\label{sec:single}

In single-stage quantization, matrix $\mbf{U}[k]$ is quantized by applying a quantization codebook $\mathcal{Q}_m^n[k]$ consisting of semi-unitary matrices $\mbf{Q} \in \mathbb{C}^{n \times m}$, $\mbfh{Q} \mbf{Q} = \mbf{I}_m$. For $b$ bits of \ac{CSI} feedback per time instant, the codebook is of size $\left|\mathcal{Q}_m^n[k] \right| = 2^b$.  

As quantization metric, we consider the subspace chordal distance, as it is relevant for many subspace-based precoding techniques, such as, block-diagonalization and interference alignment~\cite{Jindal2006,Ravindran2008,Rezaee2012,Krishnamachari2013}. The \ac{CSI} quantization problem thus is
\begin{gather}
 \hat{\mbf{U}}[k] = \argmin_{\mbf{Q} \in \mathcal{Q}_m^n[k]}\, \chord{\mbf{U}[k],\mbf{Q}}, \label{eq:quant_SS}\\
 \chord{\mbf{U}[k],\mbf{Q}} = 1 - \frac{1}{m} \trace{\mbfh{Q} \mbf{U}[k] \HH{\mbf{U}[k]} \mbf{Q} }, \label{eq:chord} 
\end{gather}
where~(\ref{eq:chord}) denotes the chordal distance normalized by the subspace dimension $m$. Solving this non-convex Grassmannian quantization problem usually implies an exhaustive search over all codebook entries, which can easily become intractable in case of large codebooks.

\subsubsection{Quantization Distortion}
In case of memoryless \ac{RVQ}, the average normalized single-stage quantization distortion is
\begin{gather}
	\mathrm{\bar{d}_{c}^2} = \EEb{\chord{\mbf{U}[k],\hat{\mbf{U}}[k]}} = \frac{1}{m} k_{n,m} 2^{-\frac{b}{m (n-m)}},
\end{gather}
with dimension-dependent constant $k_{n,m}$ as specified in~\cite{Dai2008}.

\subsubsection{Selective CSI Update}
In a temporally correlated channel, it may not be necessary to update the quantized \ac{CSI} every time instant, as the channel may not have changed sufficiently. To exploit this, we consider a simple selective \ac{CSI} update based on the quantization error w.r.t. the previously quantized \ac{CSI} 
\begin{gather}
	\hat{\mbf{U}}[k] = \begin{cases} \hat{\mbf{U}}[k-1], \text{ if } \chord{\mbf{U}[k],\hat{\mbf{U}}[k-1]} \leq c_u\, \mathrm{\bar{d}_{c}^2},\\
	\argmin_{\mbf{Q} \in \mathcal{Q}_m^n[k]}\, \chord{\mbf{U}[k],\mbf{Q}}, \text{ else.}	\label{eq:SS_update}
	\end{cases}
\end{gather} 
Here, the tuning parameter $c_u \geq 1$ determines the trade-off between the frequency of \ac{CSI} updates and the achieved average quantization distortion.

\subsection{Recursive Multi-Stage Quantization}
\label{sec:multi}
We have proposed recursive multi-stage quantization in~\cite{Schwarz_SPL20} as a means to reduce the quantization complexity in case a large quantization codebook is employed. In this approach, the \ac{CSI} is recursively quantized in $R$ stages according to
\begin{gather}
	\hmbf{U}[k] = \prod_{i = 1}^{R} \mbf{W}_i[k],\ \mbf{W}_i[k] = \hskip-5pt \argmin_{\mbf{Q} \in \mathcal{Q}_{d_i}^{d_{i-1}}[k]}\hskip-5pt \chord{\mbf{B}_{i-1}[k],\mbf{Q}},  \label{eq:quant_MS}\\
	\mbf{W}_i[k] \in \mathbb{C}^{d_{i-1} \times d_i},\ d_{i-1} > d_i,\\	
	\mbf{B}_i[k] \hskip-2pt =\hskip-2pt \HH{\mbf{W}_i[k]} \mbf{B}_{i-1}[k]\hskip-2pt \left(\HH{\mbf{B}_{i-1}[k]} \mbf{W}_i[k] \HH{\mbf{W}_i[k]} \mbf{B}_{i-1}[k] \right)^{\hskip-2pt -\frac12}\hskip-2pt, \label{eq:sqbc}
\end{gather} 
where $d_0 = n$, $d_R = m$ and $\mbf{B}_0 = \mbf{U}[k]$. Matrix $\mbf{B}_i[k] \in \mathbb{C}^{d_i \times m}$ is known as \ac{SQBC} matrix and has been derived in~\cite{Schwarz_TWC2013}. This recursive multi-stage quantizer successively reduces the dimensions of the intermediate quantizer input $\mbf{B}_i[k]$ until the intended subspace dimension $d_R = m$ is reached. 

In each of the $R$ stages of this approach, a Grassmannian quantization problem is solved. Each stage uses a quantization codebook $\mathcal{Q}_{d_i}^{d_{i-1}}[k]$ with codebook entries of dimension $d_{i-1} \times d_i$; the difference $\Delta_i = d_{i-1} - d_i$ is known as dimension step-size. Compared to single-stage quantization, however, each stage uses a much smaller codebook, since the total number of $b$ quantization bits is distributed amongst the stages, such that $\sum_{i = 1}^{R} b_i = b$. In this paper, we apply equal bit allocation $b_i = b/R$ amongst stages, even though this is suboptimal in terms of quantization distortion; yet, this choice leads to the smallest total number of codebook entries of the $R$ stages: $\sum_{i= 1}^R 2^{b_i} = \sum_{i= 1}^R 2^{b/R} = 2^{b/R}\,R \ll 2^b$. 

Furthermore, we apply a dimension step-size $\Delta_i = 1$, as this achieves the lowest quantization complexity via one-dimensional Grassmannian quantization in the orthogonal complement. Specifically, this means that instead of the minimum chordal distance quantization problem in~(\ref{eq:quant_MS}), we actually employ a quantization codebook for the one-dimensional orthogonal complements of the elements of $\mathcal{Q}_{d_i}^{d_{i-1}}[k]$, and find the codebook entry that maximizes the chordal distance. The details of this equivalent, yet less complex quantization problem formulation are explained in~\cite{Schwarz_SPL20}.     

\subsubsection{Quantization Distortion}
The average normalized chordal distance distortion of the recursive multi-stage quantizer utilizing \ac{RVQ} in each stage, a dimension step-size of $\Delta_i = 1$ and equal bit allocation is
\begin{gather}
		\mathrm{\bar{d}_{c}^2} = 1 - \prod_{i = 1}^{R} \left(1- \mathrm{\bar{d}_{c,i}^2}\right), \\
		\mathrm{\bar{d}_{c,i}^2} = \frac{1}{m} k_{d_{i-1},m,d_i} 2^{-\frac{b}{m R}}.
\end{gather}
Here, $\mathrm{\bar{d}_{c,i}^2} = \EEb{\chord{\mbf{B}_{i-1}[k],\mbf{W}_i[k]}}$ denotes the average normalized chordal distance of the $i$-th stage, which quantizes the $m$-dimensional subspace $\spans{\mbf{B}_{i-1}[k]}$ by the $d_i$-dimensional subspace $\spans{\mbf{W}_i[k]}$. The dimension-dependent constant $k_{d_{i-1},m,d_i}$ is provided in~\cite{Dai2008}.

\subsubsection{Selective Stage Update}
Similarly to single-stage quantization, we can also adopt a selective \ac{CSI} update in multi-stage quantization. Yet, here we have the additional degree of freedom to only update a subset of the stages of the quantizer, based on the currently achieved \ac{CSI} quality. Specifically, fixing the first $r$ stages of the quantizer to the previously quantized matrices $\mbf{W}_i[k-1], i \leq r$, the quantizer input matrix $\mbf{B}_{r}[k]$ of the $r+1$-th stage can be calculated recursively by replacing $\mbf{W}_i[k]$ with $\mbf{W}_i[k-1]$ in~\Cref{eq:sqbc}, and then the quantizer proceeds as usual to update the remaining stages.  

\begin{table*}[h!]  \setlength\tabcolsep{0.5em}
  \begin{center}
    \caption{Adopted neural network structure for \ac{CSI} classification of the $i$-th stage.}\vskip-5pt
    \label{tab:DNN}
    \begin{tabular}{c|c|c|c} 
      Input & 1st Layer & 2nd Layer & Output \\ \midrule
			\multirow{2}{*}{$\left[\begin{array}{c}\mathrm{real}\left(\vect{\mbf{B}_{i-1}[k]}\right) \\ \mathrm{imag}\left(\vect{\mbf{B}_{i-1}[k]}\right)\end{array}\right] $} & $15\cdot 2 d_{i-1} m$ fully connected  & $2^{b_i}$ fully connected & class output \\
			& ReLu with dropout & soft-max & cross-entropy loss 
    \end{tabular}
  \end{center}\vskip-15pt
\end{table*}

To decide how many stages the quantizer should update, we propose an approach similar to~(\ref{eq:SS_update}). If $\chord{\mbf{U}[k],\hat{\mbf{U}}[k-1]} \leq c_u\, \mathrm{\bar{d}_{c}^2}$ we do not update any stage. Otherwise, we calculate the expected distortion $\mathrm{d_{c}^2}[r']$ under the assumption that the first $r'$ stages are not updated, and determine the largest $r$ (smallest number of \ac{CSI} updates) that achieves an acceptable distortion 
\begin{gather}
	r = \argmax_{r' \in \{1,\ldots,R\}} r',\ \text{ subject to: } \mathrm{d_{c}^2}[r'] \leq c_\ell \, \mathrm{\bar{d}_{c}^2},\\
	 \mathrm{d_{c}^2}[r'] =  1 - \ldots \nonumber \\ \prod_{i = 1}^{r'} \left(1 - \chord{\mbf{B}_{i-1}[k],\mbf{W}_i[k-1]} \right) \prod_{i = r'+1}^{R} \left(1- \mathrm{\bar{d}_{c,i}^2}\right).
\end{gather}
The two tuning parameters $1 \leq c_\ell \leq c_u$ effectively define a hysteresis for the acceptable \ac{CSI} quality and thereby determine the frequency of the stage updates.

\vspace*{-0.5em}
\subsection{Deep Learning Classification}
\label{sec:NN}

The chordal distance quantization problem~(\ref{eq:quant_SS}) is essentially a classification problem and can as such, in principle, be handled by neural network structures. The problem is that the number of classes is often too large, such that a \ac{DNN} does not achieve sufficient classification accuracy. 

Consider, for example, a system with $n \times m = 8 \times 2$ antennas. If we intend to achieve an average normalized chordal distance distortion of $\mathrm{\bar{d}_{c}^2} = 0.1$, we have to employ a single-stage quantization codebook of 34\,bits, which gives an intractable number of quantization classes. In contrast, for the multi-stage quantizer, we achieve the same accuracy with 7\,bits per stage, i.e., a codebook size of 128 entries per stage, which is a number that a neural network can handle. However, the total feedback overhead is increased to 7\,bits per stage times 6 stages equals 42\,bits. Hence, for the single-stage approach~(\ref{eq:quant_SS}), the quantization problem is essentially intractable, whereas with multi-stage quantization~(\ref{eq:quant_MS}) the number of classes per stage is in fact so small that we can even adopt \ac{DNN} classification. 

As we consider Grassmannian quantization, the \ac{DNN} classification outcome should be unaffected by right-multiplication of $\mbf{B}_i[k]$ by an arbitrary unitary matrix. To exploit this invariance, we apply a phase-rotation to the individual columns of $\mbf{B}_i[k]$, such that the first row contains only real numbers, and vectorize the result before feeding into the neural network.

We summarize the adopted \ac{DNN} structure in~\Cref{tab:DNN}. We have investigated \ac{DNN} structures of varying width and depth. As can be seen from~\Cref{tab:DNN}, we employ a relatively shallow neural network with a wider first layer. Going deeper did not achieve more accurate classification. The adopted structure provides a good trade-off between complexity and achieved classification accuracy.

Training of the \acp{DNN} is achieved by generating training-sets consisting of random isotropically distributed semi-unitary matrices $\mbf{U}$, resp. $\mbf{B}_i$, and corresponding labels, obtained by solving problems~(\ref{eq:quant_SS}),~(\ref{eq:quant_MS}) through exhaustive search.   

Utilizing these \acp{DNN}, the computational complexity of the quantizer is basically off-loaded to the offline training-phase of the \acp{DNN}. The online quantization complexity is determined by the calculation of the quantizer input matrices $\mbf{B}_i[k]$~(\ref{eq:sqbc}).

\vspace*{-0.75em}
\section{Simulations}
\label{sec:Sim}

\begin{figure}
\centering
\vskip-10pt
\begin{center}
\includegraphics[width = 0.8 \columnwidth]{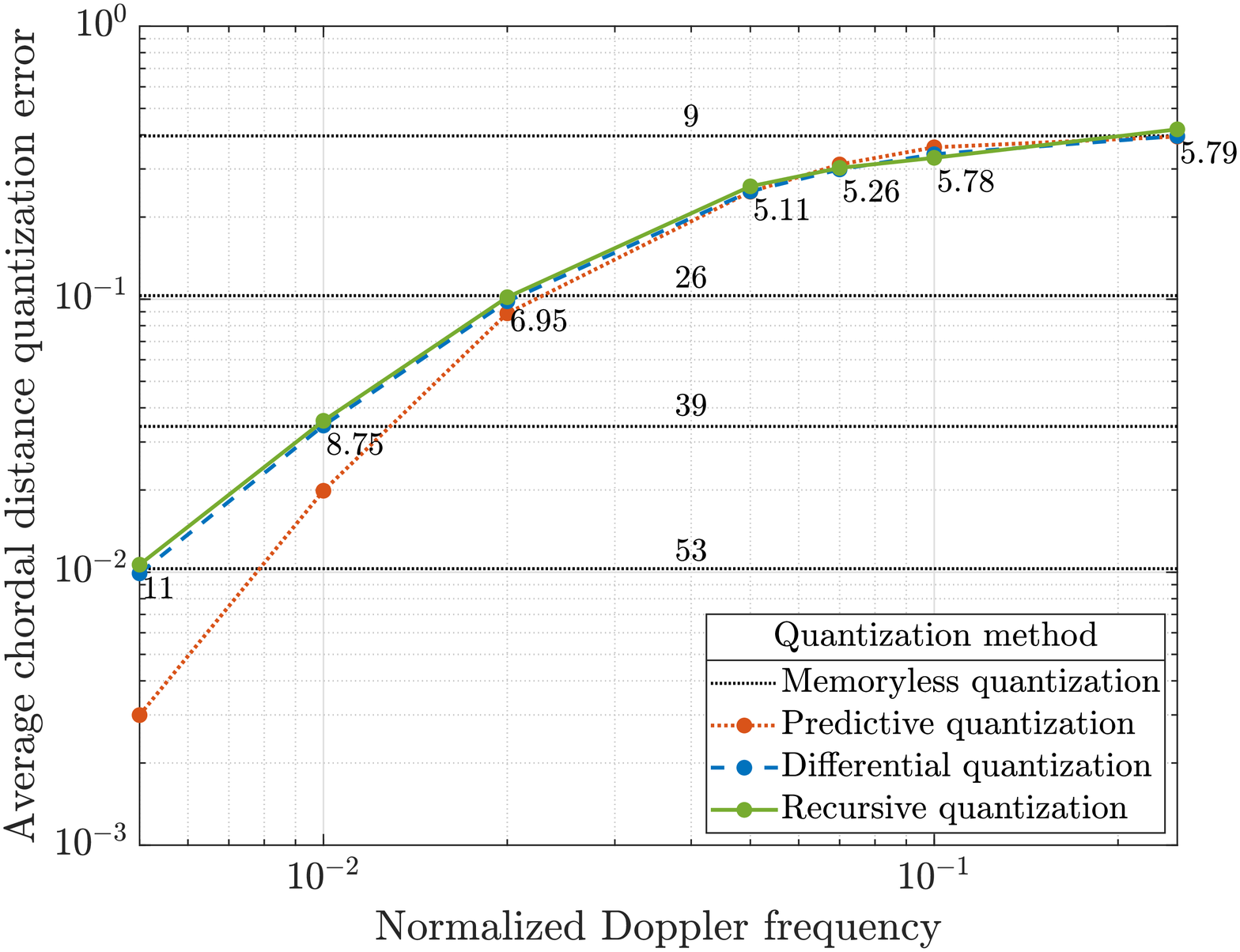} 
\vskip-5pt
\caption{Average quantization distortion on $\GG{6,2}$ versus normalized Doppler frequency. The differential and predictive quantizers provide 6\,bits of feedback per time instant; the average number of feedback bits of the recursive multi-stage quantizer is written next to the data points.}
\vskip-15pt
\label{fig:dist_6x2}
\end{center}
\end{figure}

In our first simulation, we compare recursive multi-stage quantization to differential and predictive Grassmannian quantization, utilizing the same simulation setup and quantizers as in~\cite{Schwarz_SPL_2014}. Specifically, the channel matrices $\mbf{H}[k]$ of dimension $n \times m = 6 \times 2$ are from a Rayleigh fading distribution and are temporally correlated according to Clarke's Doppler spectrum~\cite{Clarke1968} by employing the sum-of-sinusoids approach of~\cite{Zheng2003}. The predictive and differential quantizers, as proposed in~\cite{Schwarz-ACSQ2013/1}, provide 6\,bits of feedback per time instant. For comparison, we also show the performance of memoryless single-stage quantization (without selective \ac{CSI} update) for 9, 26, 39 and 53\,bits of feedback, respectively. For the recursive multi-stage quantizer, we have selected the number of quantization bits per stage to achieve the same average performance as the differential quantizer. The average feedback overhead of the recursive quantizer with selective stage update is written next to the simulated data points in~\Cref{fig:dist_6x2}. 

We observe in~\Cref{fig:dist_6x2} that at low Doppler frequencies the differential and predictive quantizers are more effective in exploiting temporal correlation than the recursive quantizer with selective stage update, as they require only 6\,bits of feedback. However, they involve on-the-fly adaptation of the entire quantization codebook at each time instant, which is hard to realize in practice. At moderate to higher Doppler frequencies all three approaches achieve very similar performance. In terms of complexity, though, the recursive quantizer is a lot more tractable, as it does not require any codebook adaptations.  

\begin{figure}
\centering
\begin{center}
\includegraphics[width = 0.8 \columnwidth]{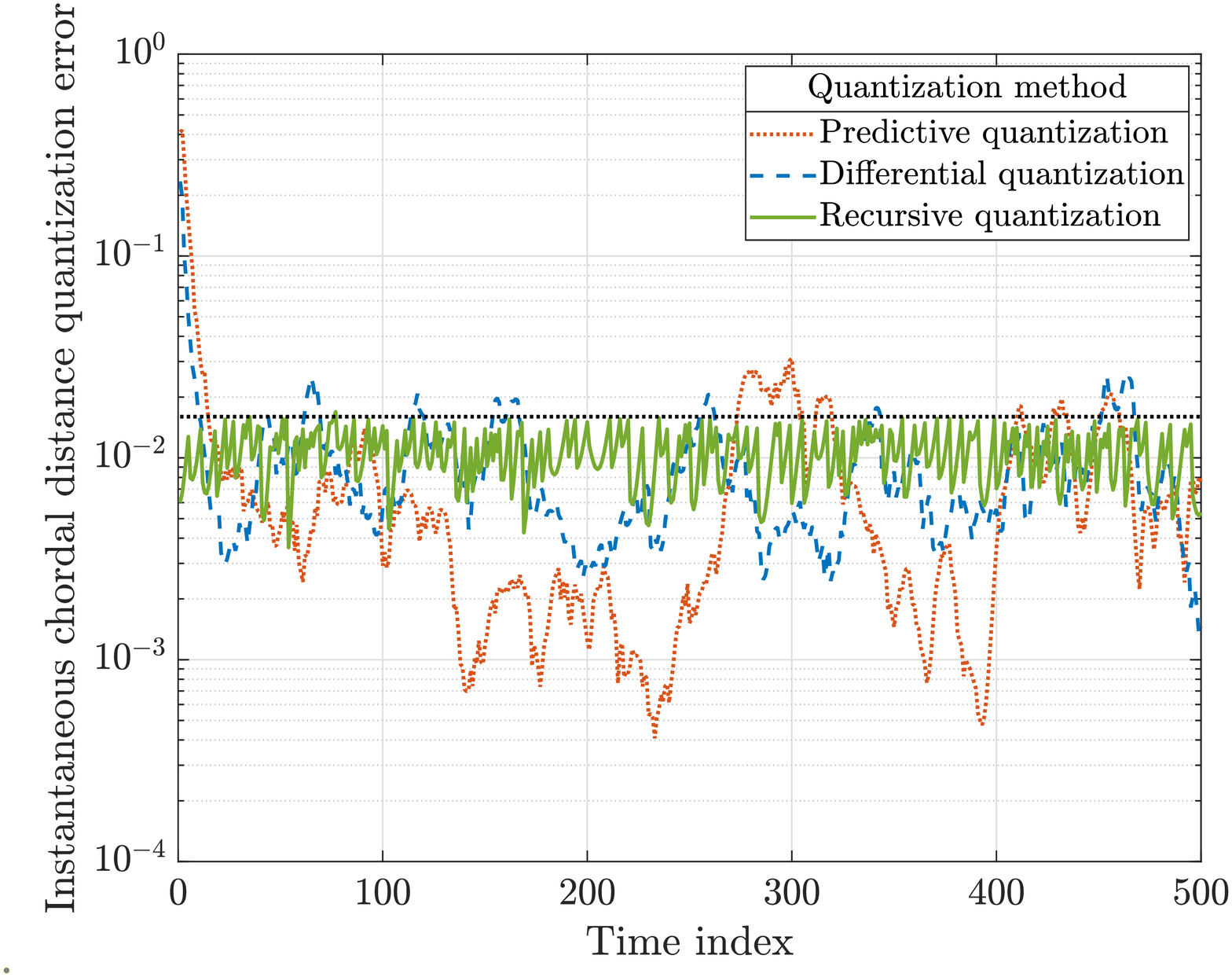} 
\vskip-5pt
\caption{Trace of the quantization error on $\GG{6,2}$ over time for the differential, predictive and recursive quantizers and a normalized Doppler of $\nu_d = 0.005$.}
\vskip-10pt
\label{fig:time_6x2}
\end{center}
\end{figure}
  
\begin{table*}[h!] \scriptsize \setlength\tabcolsep{0.5em}
  \begin{center}
    \caption{Classification and average distortion performance of the individual stages of the recursive multi-stage quantizer realized by the neural network structure of~\cref{sec:NN} for a codebook size of 6\,bits per stage.}\vskip-5pt
    \label{tab:net_perf}
    \begin{tabular}{r?c|c|c|c|c|c|c|c|c|c|c|c|c|c|c|c} 
      \textbf{Stage} & 1 & 3 & 5 & 7 & 9 & 11 & 13 & 15 & 17 & 19 & 21 & 23 & 25 & 27 & 29 & 31\\ \midrule		
			\textbf{Dimension} & 32$\times$1  & 30$\times$1 & 28$\times$1 & 26$\times$1 & 24$\times$1 & 22$\times$1 & 20$\times$1 & 18$\times$1 & 16$\times$1 & 14$\times$1 & 12$\times$1 & 10$\times$1 &  8$\times$1 &  6$\times$1 & 4$\times$1 & 2$\times$1 \\
			\textbf{Distortion} & 5.2\,e$^{-4}$ & 5.6\,e$^{-4}$ & 5.9\,e$^{-4}$ & 6.4\,e$^{-4}$ & 7.0\,e$^{-4}$ & 7.6\,e$^{-4}$ & 8.4\,e$^{-4}$ & 9.4\,e$^{-4}$ & 1.1\,e$^{-3}$ & 1.2\,e$^{-3}$ & 1.4\,e$^{-3}$ & 1.8\,e$^{-3}$ & 2.3\,e$^{-3}$ & 3.2\,e$^{-3}$ & 5.4\,e$^{-3}$ & 13.2\,e$^{-3}$ \\ \hline
			\textbf{Exhaustive} & 5.0\,e$^{-4}$ & 5.4\,e$^{-4}$ & 5.8\,e$^{-4}$ & 6.2\,e$^{-4}$ & 6.8\,e$^{-4}$ & 7.4\,e$^{-4}$ & 8.2\,e$^{-4}$ & 9.2\,e$^{-4}$ & 1.0\,e$^{-3}$ & 1.1\,e$^{-3}$ & 1.3\,e$^{-3}$ & 1.7\,e$^{-3}$ & 2.2\,e$^{-3}$ & 3.1\,e$^{-3}$ & 5.1\,e$^{-3}$ & 13.1\,e$^{-3}$
    \end{tabular}
  \end{center}\vskip-20pt
\end{table*}

In~\Cref{fig:time_6x2}, we show an exemplary trace of the instantaneous chordal distance quantization error of recursive, differential and predictive quantization for a normalized Doppler frequency of $\nu_d = 0.005$. We observe that the recursive quantizer, in contrast to the other two, has no convergence phase and exhibits relatively predictable quantization performance, which is basically dictated by the hysteresis parameters of the selective stage update. In this example, we have selected $c_u = 2$ and $c_\ell = 1.5$. The black-dotted line in the figure corresponds to the maximally acceptable distortion $c_u\, \mathrm{\bar{d}_{c}^2}$, which determines when a stage update takes place. For predictive and differential quantization, the performance shows larger variations over time, with increasing distortion whenever the quantizer cannot keep up with the temporal subspace variation of the channel.

In our second simulation, we consider quantization on $\GG{32,1}$ and employ a Gauss-Markov channel model to generate the temporally correlated channel vectors according to $\mbf{h}[k] = \alpha \mbf{h}[k -1] + \sqrt{1-\alpha^2} \mbf{g}[k]$, with \ac{iid} $\mbf{g}[k] \sim \mathcal{CN}\left(\boldsymbol{0},\mbf{I}_n\right)$ and $\alpha = J_0\left(2 \pi \nu_d \right)$. We select the codebook sizes to achieve an average chordal distance quantization distortion of $\mathrm{\bar{d}_{c}^2} = 0.06$. For single-stage quantization this requires a codebook size of 125\,bits. Obviously, we cannot implement a codebook that is this large; the results of single-stage quantization are therefore based on the theoretic performance investigations of \ac{RVQ} provided in~\cite{Dai2008}. For recursive quantization, we employ 31 stages and use a codebook size of 6\,bits per stage, i.e., a total feedback overhead of 186\,bits when all stages are updated. The individual stages of the recursive quantizer are realized by the classification neural network explained in~\Cref{sec:NN}. We summarize the normalized chordal distance distortion of the network in~\Cref{tab:net_perf}. The classification accuracy of the individual stages of the quantizer lies at approximately 90\%; however, the incurred average chordal distance distortion penalty compared to an exhaustive search is negligible.  

\begin{figure}
\centering
\begin{center}
\includegraphics[width = 0.8 \columnwidth]{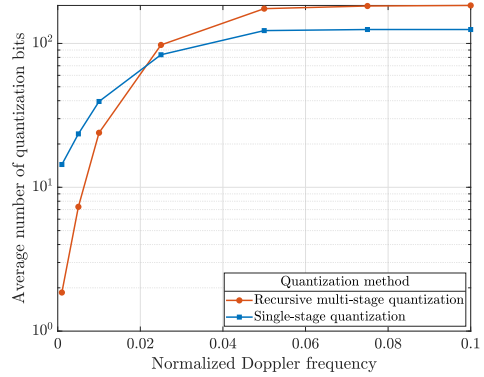} 
\vskip-5pt
\caption{Average number of quantization bits on $\GG{32,1}$ versus normalized Doppler frequency for recursive multi-stage and single-stage quantization.}
\vskip-10pt
\label{fig:bits_16x1}
\end{center}
\end{figure}

\begin{figure}
\centering
\begin{center}
\includegraphics[width = 0.8 \columnwidth]{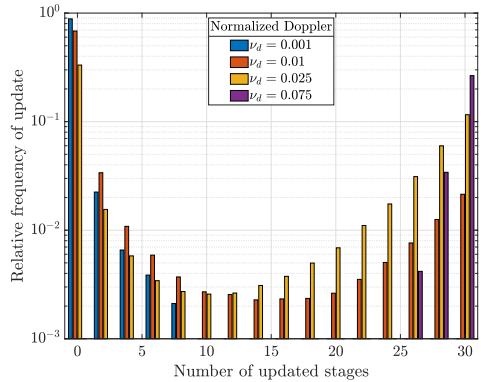} 
\vskip-5pt
\caption{Relative frequency of the number of updated stages of the recursive multi-stage quantizer for different normalized Doppler frequencies.}
\vskip-10pt
\label{fig:stages_16x1}
\end{center}
\end{figure}

In~\cref{fig:bits_16x1}, we show the average number of quantization bits of single- and multi-stage quantization with selective \ac{CSI}/stage update to achieve an average distortion of $\mathrm{\bar{d}_{c}^2} = 0.06$ as a function of the normalized Doppler frequency. We observe that at low Doppler frequencies the recursive multi-stage quantizer requires less feedback overhead than the single-stage quantizer, as it can selectively update just a subset of the stages of the quantizer. This behavior is investigated in more detail in~\cref{fig:stages_16x1}, where we plot the relative frequency of the number of updated stages of the quantizer. We can see that for small Doppler frequencies in most cases no or just few of the later stages of the quantizer are updated, whereas for high Doppler frequencies almost all stages have to be updated every time.   

\vspace*{-0.5em}
\section{Conclusion}

In this paper, we have extended recursive Grassmannian multi-stage quantization to exploit temporal channel correlations by selectively updating the individual stages of the quantizer, depending on the achieved distortion. We have shown that this approach performs similar to differential/predictive Grassmannian quantization at moderate to high Doppler frequencies, yet without requiring complicated quantization codebook adaptations. We have furthermore shown that multi-stage quantization can effectively be combined with neural network structures, as the number of classes per stage is sufficiently small to enable accurate \ac{DNN} classification. 
 
\FloatBarrier


\clearpage
\bibliographystyle{IEEEtran_no_url}
\begin{footnotesize}
\bibliography{Rec_Grass}

\begin{thebibliography}{10}
\providecommand{\url}[1]{#1}
\csname url@rmstyle\endcsname
\providecommand{\newblock}{\relax}
\providecommand{\bibinfo}[2]{#2}
\providecommand\BIBentrySTDinterwordspacing{\spaceskip=0pt\relax}
\providecommand\BIBentryALTinterwordstretchfactor{4}
\providecommand\BIBentryALTinterwordspacing{\spaceskip=\fontdimen2\font plus
\BIBentryALTinterwordstretchfactor\fontdimen3\font minus
  \fontdimen4\font\relax}
\providecommand\BIBforeignlanguage[2]{{%
\expandafter\ifx\csname l@#1\endcsname\relax
\typeout{** WARNING: IEEEtran.bst: No hyphenation pattern has been}%
\typeout{** loaded for the language `#1'. Using the pattern for}%
\typeout{** the default language instead.}%
\else
\language=\csname l@#1\endcsname
\fi
#2}}

\bibitem{Love2008}
D.~Love, R.~{Heath, Jr.}, V.~Lau, D.~Gesbert, B.~Rao, and M.~Andrews, ``An
  overview of limited feedback in wireless communication systems,'' \emph{IEEE
  Journal on Selected Areas in Communications}, vol.~26, no.~8, Oct. 2008.

\bibitem{Choi2013}
J.~Choi, Z.~Chance, D.~Love, and U.~Madhow, ``Noncoherent trellis coded
  quantization: A practical limited feedback technique for massive {MIMO}
  systems,'' \emph{IEEE Transactions on Communications}, vol.~61, no.~12, pp.
  5016--5029, December 2013.

\bibitem{Park2016}
J.~{Park} and R.~W. {Heath}, ``Multiple-antenna transmission with limited
  feedback in device-to-device networks,'' \emph{IEEE Wireless Communications
  Letters}, vol.~5, no.~2, pp. 200--203, 2016.

\bibitem{Kwon2020}
G.~{Kwon} and H.~{Park}, ``Limited feedback hybrid beamforming for multi-mode
  transmission in wideband millimeter wave channel,'' \emph{IEEE Transactions
  on Wireless Communications}, vol.~19, no.~6, pp. 4008--4022, 2020.

\bibitem{Medra15}
A.~{Medra} and T.~N. {Davidson}, ``Incremental {Grassmannian} feedback schemes
  for multi-user {MIMO} systems,'' \emph{IEEE Transactions on Signal
  Processing}, vol.~63, no.~5, pp. 1130--1143, 2015.

\bibitem{Decurninge2017}
A.~Decurninge and M.~Guillaud, ``Cube-split: Structured quantizers on the
  grassmannian of lines,'' in \emph{IEEE Wireless Communications and Networking
  Conference}, pp. 1--6, March 2017.

\bibitem{Ratnam18}
V.~V. {Ratnam}, A.~F. {Molisch}, O.~Y. {Bursalioglu}, and H.~C. {Papadopoulos},
  ``Hybrid beamforming with selection for multiuser massive {MIMO} systems,''
  \emph{IEEE Transactions on Signal Processing}, vol.~66, no.~15, pp.
  4105--4120, 2018.

\bibitem{Schwarz_JSTSP19}
S.~{Schwarz}, M.~{Rupp}, and S.~{Wesemann}, ``Grassmannian product codebooks
  for limited feedback massive {MIMO} with two-tier precoding,'' \emph{{IEEE}
  Journal of Selected Topics in Signal Processing}, vol.~13, no.~5, pp.
  1119--1135, Sep. 2019.

\bibitem{Dhillon2007}
I.~S. {Dhillon}, R.~{Heath, Jr.}, T.~{Strohmer}, and J.~A. {Tropp},
  ``Constructing packings in {G}rassmannian manifolds via alternating
  projection,'' \emph{ArXiv e-prints}, Sept. 2007.

\bibitem{Laue2017}
H.~E.~A. {Laue} and W.~P. {du Plessis}, ``A coherence-based algorithm for
  optimizing rank-1 {Grassmannian} codebooks,'' \emph{IEEE Signal Processing
  Letters}, vol.~24, no.~6, pp. 823--827, 2017.

\bibitem{Tahir_SPL19}
B.~{Tahir}, S.~{Schwarz}, and M.~{Rupp}, ``Constructing {Grassmannian} frames
  by an iterative collision-based packing,'' \emph{IEEE Signal Processing
  Letters}, vol.~26, no.~7, pp. 1056--1060, July 2019.

\bibitem{Luo2017}
X.~{Luo}, P.~{Cai}, X.~{Zhang}, D.~{Hu}, and C.~{Shen}, ``A scalable framework
  for {CSI} feedback in {FDD} massive {MIMO} via {DL} path aligning,''
  \emph{IEEE Trans. on Signal Processing}, vol.~65, no.~18, pp. 4702--4716,
  Sep. 2017.

\bibitem{Xie2017}
H.~Xie, F.~Gao, S.~Zhang, and S.~Jin, ``A unified transmission strategy for
  {TDD/FDD} massive {MIMO} systems with spatial basis expansion model,''
  \emph{{IEEE} Transactions on Vehicular Technology}, vol.~66, no.~4, pp.
  3170--3184, April 2017.

\bibitem{Schwarz_JWCN2018}
S.~Schwarz, ``Robust full-dimension {MIMO} transmission based on limited
  feedback angular-domain {CSIT},'' \emph{{EURASIP} Journal on Wireless
  Communications and Networking}, vol. 2018, no.~1, pp. 1--20, Mar 2018.

\bibitem{Wang2019}
T.~{Wang}, C.~{Wen}, S.~{Jin}, and G.~Y. {Li}, ``Deep learning-based {CSI}
  feedback approach for time-varying massive {MIMO} channels,'' \emph{IEEE
  Wireless Communications Letters}, vol.~8, no.~2, pp. 416--419, April 2019.

\bibitem{Liu2019}
Z.~{Liu}, L.~{Zhang}, and Z.~{Ding}, ``Exploiting bi-directional channel
  reciprocity in deep learning for low rate massive {MIMO} {CSI} feedback,''
  \emph{IEEE Wireless Communications Letters}, vol.~8, no.~3, pp. 889--892,
  2019.

\bibitem{Jang20}
J.~{Jang}, H.~{Lee}, S.~{Hwang}, H.~{Ren}, and I.~{Lee}, ``Deep learning-based
  limited feedback designs for {MIMO} systems,'' \emph{IEEE Wireless
  Communications Letters}, vol.~9, no.~4, pp. 558--561, 2020.

\bibitem{Inoue2009}
T.~Inoue and R.~{Heath, Jr.}, ``Grassmannian predictive coding for delayed
  limited feedback {MIMO} systems,'' in \emph{47th Annual Allerton Conference
  on Communication, Control, and Computing}, Oct. 2009.

\bibitem{Sacristan2010}
D.~Sacristan-Murga and A.~Pascual-Iserte, ``Differential feedback of {MIMO}
  channel {G}ram matrices based on geodesic curves,'' \emph{IEEE Trans. on
  Wireless Communications}, vol.~9, no.~12, pp. 3714--3727, Dec. 2010.

\bibitem{Ayach2012}
O.~El~Ayach and R.~{Heath, Jr.}, ``Grassmannian differential limited feedback
  for interference alignment,'' \emph{IEEE Transactions on Signal Processing},
  vol.~60, no.~12, pp. 6481--6494, Dec 2012.

\bibitem{Schwarz-ACSQ2013/1}
S.~Schwarz, R.~{Heath, Jr.}, and M.~Rupp, ``Adaptive quantization on the
  {G}rassmann-manifold for limited feedback multi-user {MIMO} systems,'' in
  \emph{38th International Conference on Acoustics, Speech and Signal
  Processing}, pp. 5021 -- 5025, Vancouver, Canada, May 2013.

\bibitem{Schwarz_SPL_2014}
S.~Schwarz and M.~Rupp, ``Predictive quantization on the {S}tiefel manifold,''
  \emph{{IEEE} Signal Processing Letters}, vol.~22, no.~2, pp. 234--238, 2015.

\bibitem{Ge2018}
Y.~{Ge}, Z.~{Zeng}, T.~{Zhang}, and Y.~{Liu}, ``Spatio-temporal correlated
  channel feedback for massive {MIMO} systems,'' in \emph{{IEEE/CIC}
  International Conference on Communications in China}, pp. 1--5, 2018.

\bibitem{Lu2019}
C.~{Lu}, W.~{Xu}, H.~{Shen}, J.~{Zhu}, and K.~{Wang}, ``Mimo channel
  information feedback using deep recurrent network,'' \emph{IEEE
  Communications Letters}, vol.~23, no.~1, pp. 188--191, 2019.

\bibitem{Li2020}
X.~{Li} and H.~{Wu}, ``Spatio-temporal representation with deep neural
  recurrent network in mimo csi feedback,'' \emph{IEEE Wireless Communications
  Letters}, vol.~9, no.~5, pp. 653--657, 2020.

\bibitem{Schwarz_SPL20}
S.~{Schwarz} and M.~{Rupp}, ``Reduced complexity recursive grassmannian
  quantization,'' \emph{IEEE Signal Processing Letters}, vol.~27, pp. 321--325,
  2020.

\bibitem{Clarke1968}
R.~H. Clarke, ``{A statistical theory of mobile radio reception},'' \emph{Bell
  Systems Technical Journal}, vol.~47, pp. 957--1000, 1968.

\bibitem{Jindal2006}
N.~Jindal, ``{MIMO} broadcast channels with finite-rate feedback,'' \emph{IEEE
  Transactions on Information Theory}, vol.~52, no.~11, p.~5, Nov. 2006.

\bibitem{Ravindran2008}
N.~Ravindran and N.~Jindal, ``Limited feedback-based block diagonalization for
  the {MIMO} broadcast channel,'' \emph{IEEE Journal on Selected Areas in
  Communications}, vol.~26, no.~8, pp. 1473 --1482, Oct. 2008.

\bibitem{Rezaee2012}
M.~Rezaee and M.~Guillaud, ``Limited feedback for interference alignment in the
  {K}-user {MIMO} interference channel,'' in \emph{Proc. Information Theory
  Workshop}, pp. 1--5, Lausanne, Suisse, September 2012.

\bibitem{Krishnamachari2013}
R.~Krishnamachari and M.~Varanasi, ``Interference alignment under limited
  feedback for {MIMO} interference channels,'' \emph{IEEE Transactions on
  Signal Processing}, vol.~61, no.~15, pp. 3908--3917, Aug 2013.

\bibitem{Dai2008}
W.~Dai, Y.~Liu, and B.~Rider, ``Quantization bounds on {G}rassmann manifolds
  and applications to {MIMO} communications,'' \emph{IEEE Trans. on Information
  Theory}, vol.~54, no.~3, pp. 1108 --1123, March 2008.

\bibitem{Schwarz_TWC2013}
S.~Schwarz and M.~Rupp, ``Subspace quantization based combining for limited
  feedback block-diagonalization,'' \emph{{IEEE} Transactions on Wireless
  Communications}, vol.~12, no.~11, pp. 5868--5879, 2013.

\bibitem{Zheng2003}
Y.~R. Zheng and C.~Xiao, ``Simulation models with correct statistical
  properties for {R}ayleigh fading channels,'' \emph{IEEE Transactions on
  Communications}, vol.~51, no.~6, pp. 920 -- 928, June 2003.

\end{thebibliography}
\end{footnotesize}

\end{document}